\documentclass[aps,pre,floats,floatfix,twocolumn,fleqn]{revtex4-2}

\usepackage{graphicx}
\usepackage{epstopdf}
\graphicspath{{./Figs/}} 

\usepackage{latexsym}
\usepackage{amsmath}
\usepackage{amssymb}
\usepackage{bm}
\usepackage{wasysym}

\usepackage{ifthen}
\usepackage{color}
\usepackage[colorlinks,allcolors=blue,bookmarks=true]{hyperref}

\newcommand{\mass}{\mathsf{m}}

\newcommand{\braket}[1]{\left\langle #1 \right\rangle }

\newcommand{\beq}{\begin{eqnarray}}
\newcommand{\eeq}{\end{eqnarray}}

\newcommand{\hide}[1]{}  
\newcommand{\rmrk}[1]{#1} 

\newcommand{\Eq}[1]{{\textcolor{blue}{Eq.}}~\!\!(\ref{#1})} 
\newcommand{\Sec}[1]{{\textcolor{blue}{Sec.}}~(\ref{#1})} 
\newcommand{\App}[1]{{\textcolor{blue}{Appendix}}~\ref{#1}} 
\newcommand{\Fig}[1] {{\textcolor{blue}{Fig.}}~\!\!\ref{#1}}
\newcommand{\sect}[1]{{\bf #1.-- }}

\newcommand{\hrefl}[2]{\href{#2}{(#1)}}

\begin{document}

\title{Stochastic modelling of \rmrk{spreading and dissipation} \\ in mixed-chaotic system that are driven quasi-statically} 

\author{Yehoshua Winsten and Doron Cohen} 

\affiliation{
\mbox{Department of Physics, Ben-Gurion University of the Negev, Beer-Sheva 84105, Israel} 
}

\begin{abstract}
We analyze \rmrk{energy spreading} for a system that features mixed chaotic phase-space,  
whose control parameters (or slow degrees of freedom) vary quasi-statically. 
For demonstration purpose we consider the restricted 3~body problem, 
where the distance between the two central stars is modulated due to their Kepler motion. 
If the system featured hard-chaos, one would expect diffusive spreading 
with coefficient that can be estimated using linear-response (Kubo) theory. 
But for mixed phase space the chaotic sea is multi-layered.    
Consequently, it becomes a challenge to find a robust procedure that translates  
the sticky dynamics into a stochastic model.
We propose a Poincaré-sequencing method  
that reduces the multi-dimensional motion 
into a one-dimensional random-walk in impact-space.
We test the implied relation between stickiness 
and the \rmrk{rate of spreading}.
\end{abstract}

\maketitle

\section{Introduction}

Considering a closed Hamiltonian driven system, such as a particle in a box with moving wall (the piston paradigm), the textbook assumption is that quasi-static processes are adiabatic, and therefore  {\em reversible}. This claim can be established for an {\em integrable} system by recognizing that the action-variables are adiabatic invariants~\cite{Landau}. At the opposite extreme, analysis of slowly driven completely {\em chaotic} systems \cite{Ott1,Ott2,Ott3} has led to a mesoscopic version of Kubo linear-response theory and its associated fluctuation-dissipation phenomenology \cite{Wilkinson1,Wilkinson2,crs,frc}. 
However, generic systems are neither integrable nor completely chaotic. 
Rather their phase space is {\em mixed}, 
resulting in the failure of the adiabatic picture \cite{Kedar1,Kedar2,Kedar3,apc,lbt}, 
and of linear-response theory. 
Namely, the phase space structure varies with the control parameter: 
tori are destroyed; 
chaotic corridors are opened allowing migration between different regions in phase space \cite{bhm,qtp}; 
stochastic regions merge into chaos; 
sticky regions are formed \cite{stk,stk0,stk1,stk2,stk3};  
sets of tori re-appear or emerge. 
Some of those issues can be regarded as a higher-dimensional version 
of non-linear scenarios that are relate to bifurcations of fixed points, 
notably swallow-tail loops \cite{Swallow1,Swallow2,Swallow3,Swallow4,Swallow5},
or as a higher-dimensional version of the well-studied separatrix crossing   \cite{Kruskal,Neishtadt1,Timofeev,Henrard,Tennyson,Hannay,Cary,Neishtadt2,Elskens,Anglin,Neishtadt3,lbd}, 
where the Kruskal-Neishtadt-Henrard theorem is followed.

\sect{\rmrk{Motivation}}
The analysis of driven systems that feature an underlying mixed-chaotic phase-space is a rather universal theme, that has relevance to many fields in physics. 
There are mainly two ways to motivate the quasi-static perspective for the pertinent degrees of freedom (dof). For some systems it is natural to distinguish between slow (`heavy') dof, and fast (`light') dof. Then it makes sense to regard the heavy dof as parametric driving, and to ignore the back reaction. The heavy dof might be the location of a piston, or it might be the distance between the two stars that perform Kepler motion in the restricted 3~body problem (which we discuss below).

A different way to motivate this perspective originates from mesoscopic physics. One would like to provide a comprehensive set of tools for the design and for the optimization of quasi-static protocols, e.g. in the context of Bose-Hubbard systems  \cite{exprBHH1,exprBHH2,KolovskyReview}. The feasibility and the efficiency of such protocols is related to the underlying mixed-chaotic phase-space dynamics, as demonstrated in \cite{apc,qtp,lbt}.

\hide{
The Bose-Hubbard Hamiltonian introduces an over-complicated challenging arena for demonstration of anomalies in the quasi-static limit. Such anomalies arise because this Hamiltonian features an underlying mixed-chaotic phase-space.
Recently, we have studied the efficiency of a nonlinear stimulated Raman adiabatic passage (STIRAP) \cite{apc}; the efficiency of quasi-static transfer protocols \cite{qtp}; and the Hamiltonian hysteresis that follows the reversal of the driving \cite{lbt}. 
Irreversibility is observed in hysteresis experiments with ultracold atoms \cite{exprDimerHys}. 
For ring geometry see \cite{exprRingRev,exprRingNIST}. 
Signatures of mixed phase space naturally arise in the analysis 
of the Bose-Hubbard Hamiltonian \cite{exprBHH1,exprBHH2,KolovskyReview}. 
}

\sect{\rmrk{Model systems}}
The simplest way to demonstrates anomalies that may arise in the quasi-static limit is to study billiard systems \cite{Kedar1,Kedar3}. The geometric construction allows a sharp distinction between regions in (phase)space. For example, the Bunimovich mushroom geometry of \cite{Kedar3} is composed of a regular region (the mushroom cap), and a chaotic region (stadium-like stem). However, such model is in some sense not generic. More generally, phase space has hierarchical structure with peripheral sticky regions \cite{stk} \rmrk{(and see Refs.[1-10] therein)}; and the composition of the energy surface depends on energy. Furthermore, in practice the distinction between the ``sea" and the ``islands" is not sharp neither fully controlled. This requires the development of new tools, to facilitate the analysis of the time-dependent dynamics. 

For the purpose of developing tools for the analysis of quasi-static scenarios, Billiards are too simple, while Bose-Hubbard systems are over-complicated and too demanding. A mathematically-oriented strategy would be to select an artificial Hamiltonian. But it is much more appealing to consider a toy Hamiltonian that has physical significance. At this point, it is appropriate to recall that the discussion of Hamiltonian chaos is historically rooted in the 3~body problem of celestial mechanics.

\sect{\rmrk{The restricted 3~body problem}}
It is natural to select Hill's Hamiltonian \cite{hill1,hill2,hill3} as a prototype model for analysis. This Hamiltonian describes the motion of a test-particle in the field of force of a binary systems (stars that perform Kepler motion). In reality the test particle might be a satellite or a circumbinary planet \cite{hill7,hill8,hill9}.  Optionally, in order to emphasize the analogy with the piston paradigm, we can have in mind a binary system immersed in a cloud of dust: the dust is driven quasi-statically by the Kepler motion of the stars. In reality the `dust' might be an asteroid system, and the binary system might consist of massive black holes at the center of a galactic nuclei.   

Hill's Hamiltonian, unlike the Bose-Hubbard Hamiltonian, is simple for visualization, and still possesses all the generic features of realistic models. The test-particle might perform quasi-regular motion around one of the stars, or chaotic motion wandering between the two stars. We find, as expected, a textured phase space structure with sticky peripheral regions. As an additional bonus this model also allows to consider a disintegration scenario: the test particles gains energy and eventually escapes to infinity.

\sect{\rmrk{The full 3~body problem}}
The analysis of the Hill's Hamiltonian has possibly importance in the restricted sense, but we would like to suggest that its quasi-static perspective might be of interest also for the full 3~body problem. Here we would like to refer to the recent works of \cite{NStone,HPerets} \rmrk{(and see references therein)}. Given the total energy and the total angular momentum, the challenge is to calculate, say, the probability $\sigma(E)$ that one of the stars is ejected with an energy~$E$. For this purpose it has been assumed that phase space is composed of a totally chaotic {\em interaction region}, and an outer region where one of the bodies becomes an outsider (possibly unbounded).  Assuming that the motion in the interaction region completely ergodizes the energy, the probability $\sigma(E)$, up to normalization, is given by the corresponding phase-space volume, that can be calculated analytically. 

Let us speculate that in some cases the dynamics of the escaping body, while in the interaction region, is described by the Hill's Hamiltonian. Then the question arises, how its energy~$E$ is affected by the motion of the binary system. It is possibly more transparent to re-phrase this questions using the language of statistical mechanics. Namely, considering a cloud of trajectories, we would like to analyse the {\em spreading} in energy. 

Clearly the assumption of total randomization of $E$ is an over-simplification for several reasons. First of all, integrable islands should be excluded. But even if we ignore the islands, we are going to show that the spreading dynamics is not trivial. Roughly speaking we are going to characterize the energy distribution by its ``width", and by its ``average". One expects a fluctuation-dissipation relation that related the rate of energy increase to the rate of the spreading \cite{Wilkinson1,Wilkinson2,crs,frc}. But this relation is endangered by the mixed phase space dynamics.

\sect{Outline}
In \Sec{sec:H} we introduce the generalized Hill's Hamiltonian. 
This Hamiltonian will be used as a test case for the application of our approach. 
It features a mixed chaotic phase space 
whose parametric evolution can be 
visualized using a Poincaré landscape plot.
In \Sec{sec:PSM} we use a Poincaré-sequencing method 
in order to encode the time dependent dynamics.
Consequently, the multi-dimensional motion in phase space 
is reduced into a one-dimensional random-walk in impact-space.
This inspires the introduction of an effective stochastic model   
in \Sec{sec:model1} and \Sec{sec:model2},  
which is used in \Sec{sec:SnE} to provide an explicit relation between 
stickiness and the rate of \rmrk{spreading}. 
In \Sec{sec:cycle} we explain that the dependence on the directionality 
of a cycle is linked to asymmetry that can be detected 
in the Poincaré-sequencing analysis. 
For completeness we present in \Sec{sec:dissipation} the theoretical 
reasoning that relates the rate of dissipation to 
the stochastic characterisation of the dynamics.

\section{The generalized Hill problem} \label{sec:H}

The Hamiltonian under consideration concerns the motion of a test particle (satellite) in the vicinity of massive bodies (stars). The stars are performing a cycle $(X(t),Y(t))$ that has frequency $\Omega$ and constant angular momentum~$\ell$. It might be, but not have to be, the Kepler motion of \App{app:kepler}. We define the characteristic radius~$c$ such that the scaled angular momentum is ${\ell \equiv c^2\Omega}$. In polar coordinates the cycle is parameterized by ${ R(t) = c\mathsf{R}(\theta(t)) }$. By definition of~$c$, one observes that the $d\theta/(2\pi)$ integral over $|\mathsf{R}(\theta)|^2$ is unity. Regarding $\theta$ as the time variable, one obtains, after a sequence of transformations (see \App{app:hillG}), the generalized Hill's Hamiltonian:
\beq \label{eq:H}
\mathcal{H} =  
\frac{1}{2}(\bm{p} - \bm{r}_{\perp} )^2 + g \mathsf{R}(\theta) u(\bm{r}) - \frac{1}{2} \mathsf{K}(\theta) \bm{r}^2  
\ \ \ \ \ 
\eeq        
where (prime indicates theta derivative):
\beq \label{eq:K}
\mathsf{K}(\theta) \ = \ 1 +  \left(\frac{1}{\mathsf{R}(\theta)}\right)'' \mathsf{R}(\theta)
\eeq
and the scaled version of the attractive potential is 
\beq
u(\bm{r}) = -\frac{\mu_2}{\sqrt{(x-\mu_1)^2+y^2}} -\frac{\mu_1}{\sqrt{(x+\mu_2)^2+y^2}} 
\ \ \ \ 
\eeq
with ${\mu_1+\mu_2=1}$. The parameter~$g$ is the scaled attraction constant for the force between the satellite and the stars. It can be due to gravitation, or (in different context) it can be of Coulomb origin. 

For an arbitrary quasi-Kepler motion (as defined above, meaning that $\ell$ is constant) the Hamiltonian is controlled by two parameters $(\mathsf{R},\mathsf{K})$. So in general the satellite experiences a {\em cycle}. But for a proper Kepler motion 
${\mathsf{K}(\theta) =  g_{\varepsilon} \mathsf{R}(\theta) = 1/[1+\varepsilon\cos(\theta)] }$ 
with ${g_{\varepsilon} = (1-\varepsilon^2)^{-3/4}}$. 
Consequently, see \App{app:hillK}, we get a Hamiltonian that depends on a single parameter,   
\beq 
\mathcal{H}(\bm{r},\bm{p}) 
= \frac{1}{2}(\bm{p} - \bm{r}_{\perp} )^2 +  \mathsf{R}(\theta) \left( g u(\bm{r}) - \frac{1}{2}g_{\varepsilon} \bm{r}^2 \right) 
\ \ \ \  
\eeq        
Thus, a proper Kepler motion should be regarded as a {\em modulation} and not as a {\em cycle}. 

In the last paragraph of \App{app:hillK} we explain that the dimensionless slowness parameter that indicates a quasi-static Kepler driving is ${ \varepsilon g_{\varepsilon} / g }$. 
For simulations we used $g{=}25$, and $\varepsilon{=}0.2$, and $\mu_1 {=} \mu_2 {=} 1/2$.

\begin{figure}
\includegraphics[width=8.7cm]{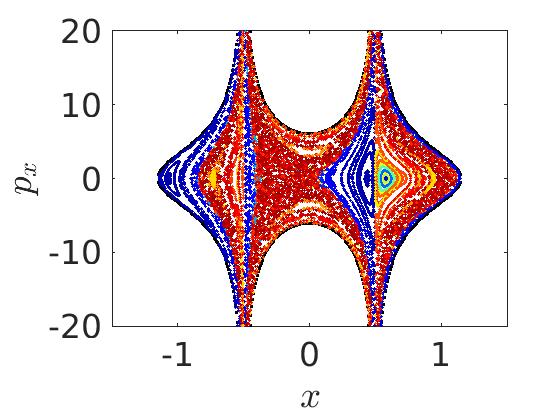}
\includegraphics[width=8.7cm]{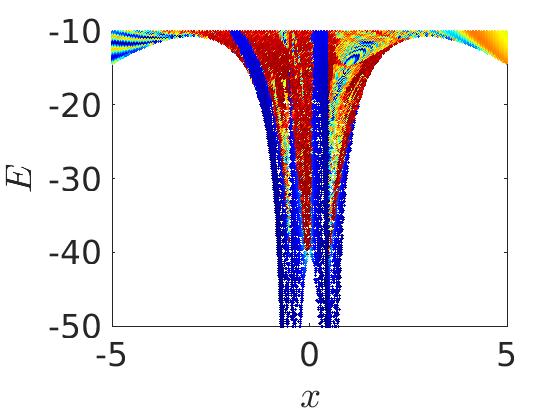}
\includegraphics[width=8.7cm]{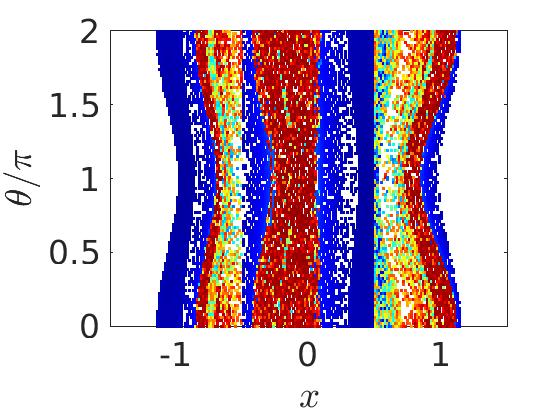}

\caption{
{\bf Poincaré landscape.}
The upper panel is a standard Poincaré section for Hill's Hamiltonian \Eq{eq:H} 
with ${\mu_1{=}\mu_2{=}1/2}$, 
and $\varepsilon{=}0.2$, 
and $g{=}25$, 
at energy~$E{=}-22.2$.  
The control parameter $\mathsf{R}$ is frozen at the value~${\theta=0}$. 
In the middle panel each row is a $p_x{=}0$ stripe of the Poincaré section at different~$E$,
while $\mathsf{R}$ is frozen at the value~$\theta{=}0$. 
In the lower panel the Poincaré stripes are plotted 
for frozen $\mathsf{R}$ at different values of~$\theta$. 
Initially~$E{=}-22.2$, and later we follow $E$ adiabatically.    
The color code is such that chaotic trajectories are red, 
while quasi-regular chaotic trajectories are blue. 
} 
\label{fLandscape}
\end{figure}

\sect{Poincaré landscape}
\Fig{fLandscape} displays a representative Poincaré section for the \rmrk{time-independent ($\theta$-frozen)} Hill's Hamiltonian. The phase space structure is as follows: two (blue) regions contain quasi-regular trajectories around each of the two stars; there are additional quasi-regular regions; and there is a large (red) chaotic sea. In order to demonstrate the variation of phase space with respect to the energy~$E$, or with respect to a control parameter (here it is $\theta$ that parameterize the Kepler motion), we propose to look on the {\em Poincaré landscape} that is displayed in the additional panels of \Fig{fLandscape}. Each \rmrk{row} in those additional panels encodes the information regarding the phase-space structure for a different value of~$E$ or~$\theta$, respectively. 

\rmrk{In a later section we display on top of this landscape, an evolving cloud that is propagated by the time-depended Hamiltonian $\mathcal{H}(\bm{r},\bm{p}; \theta(t))$, with $\dot{\theta}{=}1$ as implied by our definitions of scaled time. In this time dependent scenario, points of the cloud can spread in energy, and migrate between different regions.}

\begin{figure}
\includegraphics[width=8.5cm]{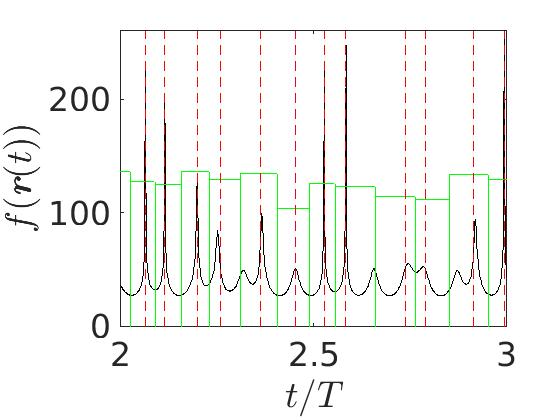}

\caption{
{\bf Poincaré pulses.}
Representative piece of the signal ${f(\bm{r}(t))}$. 
The vertical red lines indicate the moments $t_j$ that are selected by the Poincaré section.
The signal is regarded as a sequence of rectangular pulses (green line). 
Each pulse has duration $T_j$ and average height ${\bar{F}_j = F_j/T_j }$. 
In the figure $\bar{F}$ is scaled vertically ($\times 3$) to improve resolution.    
} 
\label{fPulse}  
\end{figure}

\section{Poincaré sequencing} 
\label{sec:PSM}

The spreading of energy \rmrk{of a driven system}  
is determined by the fluctuations of the generalized force~$\mathcal{F}$ 
that is associated with the control variable~$\theta$. 
Note that we assume periodic driving, 
and that the scaled Hamiltonian is defined such that $\dot{\theta}{=}1$. 
For typical model systems, e.g. Billiards with moving piston, 
and also for the Hill's Hamiltonian, we can factorize $\mathcal{F}$ as follows:    
\beq \label{eq:E}
\mathcal{F} \ \ = \ \ -\frac{\partial \mathcal{H}} {\partial \theta} \ \ \equiv \ \ \rmrk{h(\theta)} f(\bm{r})
\eeq
\rmrk{where ${f(r) = (g/g_{\varepsilon})u(r)-(1/2)r^2}$.}
The variation of the energy is an integral over $-\dot{\theta} \mathcal{F}(t)$, 
but for the analysis it is more convenient to consider 
\beq \label{eq:Q}
Q \ \ = \ \ \int_0^t f(\bm{r}(t)) dt  \ \ \equiv  \ \ \sum_j F_j
\eeq
The last equality expresses the integral as a sum over pulses, 
whose area is defined in the illustration of \Fig{fPulse}.   

\rmrk{The variation of $Q$, unlike that of~$E$ does not include $\dot{\theta}$ as a prefactor, and therefore allows, on equal footing, to compare the fluctuations of the driven system to the fluctuations that are generated by the time independent (frozen $\theta$) Hamiltonian.}


\begin{figure}
\includegraphics[width=7cm]{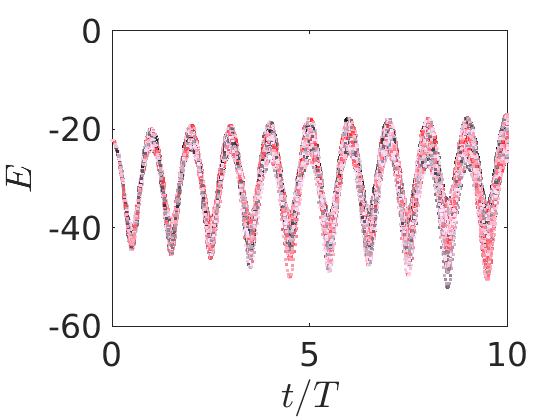} 
\includegraphics[width=7cm]{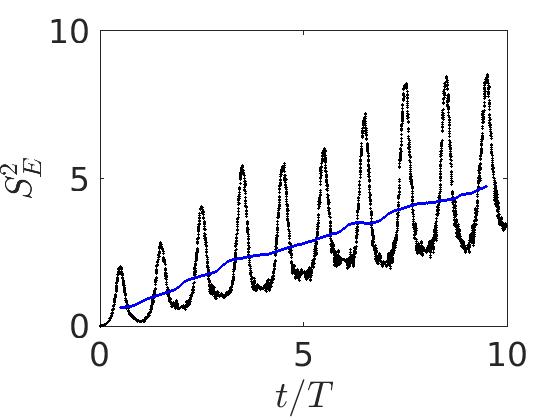} 
\caption{
{\bf Variation of the energy}. 
The energy for a cloud of chaotic trajectories is plotted as a function of time (upper panel).
The spreading is displayed in the lower panel (black line).  
Here we define $S_E$ as the width of the central region that supports 50\% of the distribution.  
The blue line is a moving average. 
} 
\label{fig:energy}  
\end{figure}

\begin{figure}
\includegraphics[width=8cm]{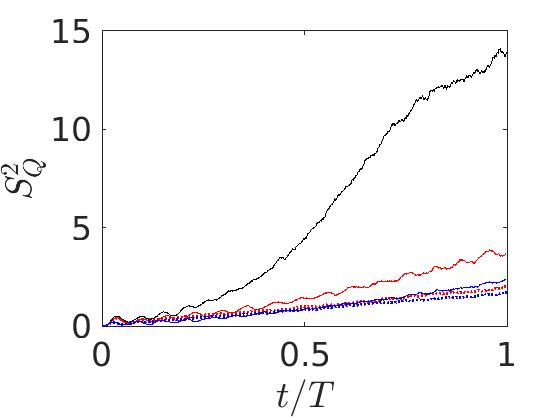}  
\includegraphics[width=8cm]{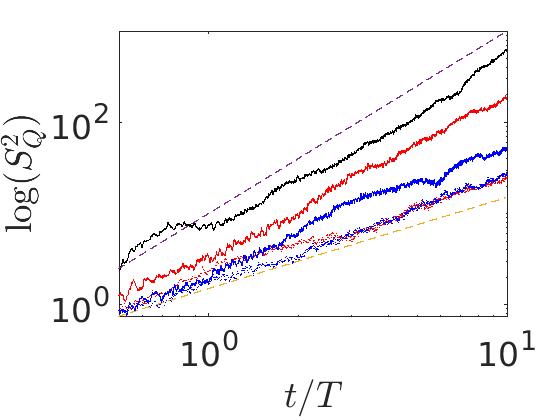}  
\caption{
{\bf \rmrk{Spreading}.}
The dynamics is generated by \Eq{eq:H}, 
with the same parameters as in \Fig{fLandscape}. 
The time dependence of $S_Q$ is regarded 
as a measure for \rmrk{spreading}, 
where $Q$ is defined by \Eq{eq:Q}.  
The black line is $S_Q$ for an ensemble of 1-cycle (normal panel) 
and 10-cycles (log-log panel) trajectories that were launched in the chaotic sea.  
In the log-scale panel, the lower and upper dashed lines indicate $\propto t$  and $\propto t^2$ dependence. 
The red and blue lines are $S_Q$ if the control parameter~$\mathsf{R}$
were frozen. The selected values are at $\theta{=}0$ (red) and $\theta{=}\pi$ (blue), 
for which the spreading is relatively fast/slow respectively.   
According to the traditional paradigm of quasi-static processes, 
the black line should be roughly between the red and the blue lines, which is clearly not the case.   
The red and blue dotted lines (lower data lines in both panels) 
are $S_Q$ for a signal that is composed of a randomized 
sequence of the same pulses, i.e. they provide the variance that would be expected 
if the actual signal did not have phase-space correlations.   
We conclude that correlations are {\em enhanced} due to the time dependence of~$\theta(t)$. 
} 
\label{fSpreading}
\end{figure}

\sect{\rmrk{Simulations}}
In the simulations we consider the following scenario. 
Initially we launch a narrow cloud in the middle of the chaotic sea.
\rmrk{After a short transient, keeping $\theta$ frozen,} 
this cloud fills most of the chaotic sea. 
Some extra time might be required in order to penetrate into
peripheral regions where the dynamics is sticky. 
We shall come back to this stickiness issue later on. 
\rmrk{Subsequently, we run the simulation with the time-dependent Hamiltonian ($\theta$ unfrozen). 
Due to the driving, the cloud further evolves as follows:} 
{\bf (a)}~Spreading away from the initial energy surface; 
{\bf (b)}~Migration between separated phase space regions. 
The dissipation aspect (growth of the average-energy) 
is directly related  to \#a and indirectly related to \#b.

\sect{\rmrk{The spreading measure $S_Q$}}
The traditional measure for phase space spreading is \rmrk{entropy},    
but we prefer to adopt a measure that has a direct practical meaning.  
\rmrk{The natural choice is to look on the energy. 
We define~$S_E$ as the width of the energy distribution, 
namely, it is the range around the median  
where $50\%$ of the distribution is located (third quartile minus first quartile).
In \Fig{fig:energy} we display~$E$ for the trajectories of the cloud,  
and extract the spreading $S_E$ as a function of time.  
Both feature intra-cycle modulation that is mainly related to the $h(\theta)$ of \Eq{eq:E}.} 
In order to get rid of this modulation we prefer to look on~$Q$. 
In \Fig{fSpreading} we display $S_Q$ as a function of time.  
It is defined as the width that holds $50\%$ of the distribution. 
Its variation in time, unlike that of $S_E$ is rather smooth,  
and better reflects the systematic spreading of the distribution 
over phase-space cells. 
\rmrk{Another advantage is that we can compare the $S_Q$ of the driven system 
with the $S_Q$ of the forozen-$\theta$ Hamiltonian. 
Clearly in the latter context $S_Q$ is not a measure for spreading, 
but a measure for the fluctuations of $\mathcal{F}$. }

\begin{figure}[b]
\includegraphics[width=6cm]{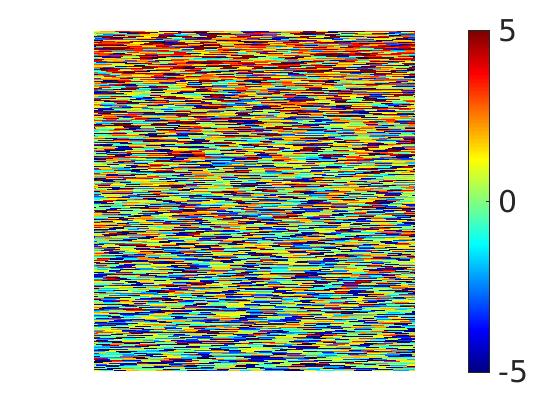}
\includegraphics[width=6cm]{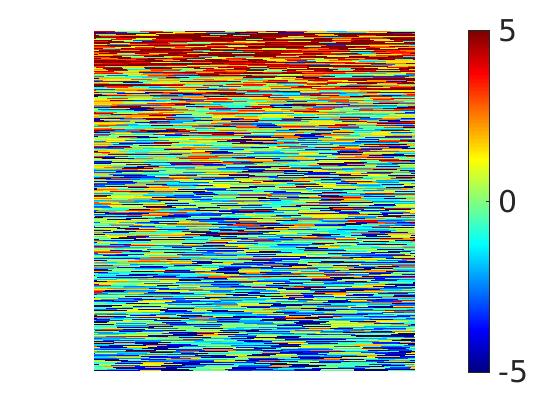}
\includegraphics[width=6cm]{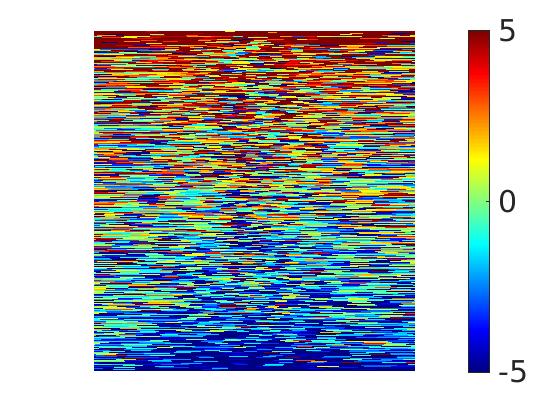}

\caption{
{\bf Poincaré sequences of pulses.}
The rows of each image displays the color-coded sequences $\bar{F}_j$ 
of the length-$T$ trajectories for which the spreading has been calculated in the upper panel of \Fig{fSpreading}.
The 3 panels are for $\theta{=}\pi$, and for $\theta{=}0$, and for the actual $\theta(t)$ of the Kepler motion.  
The trajectories in each panel are ordered by the average value of $\bar{F}_j$. 
The {\em red stretches} for $\theta{=}0$ indicate stickiness in red regions 
that will be identified in \Fig{fPoincare}.
The additional {\em blue stretches} in the lower panel 
indicate excess dwell time in the chaotic sea.
} 
\label{fPulseSeq}  
\end{figure}

\sect{\rmrk{Optional perspective on $S_Q$}}
A very long chaotic trajectory that explores the whole chaotic sea 
can be regarded as a {\em Poincaré sequence} of $F_j$ pulses 
\rmrk{that is characterized by the~$S_Q(t)$ of \Fig{fSpreading}.}    
In order to get $S_Q(t)$, the chaotic trajectory can be divided 
into sub-sequences of length~$T$ (upper panel) 
or of length~$10T$ (lower panel), 
where~$T$ is the period of a cycle.
Equivalently, \rmrk{as described in the previous paragrpah}, 
we start the simulation with a cloud of initial points 
at the middle of the chaotic sea, evolve them, 
and care to exclude the initial transient.

\sect{Detecting correlations}
In order to figure out whether temporal correlations are important 
we randomize the original $F_j$ sequence, 
and then divide it again into sub-sequences. 
In \Fig{fSpreading}, the spreading of the randomized-trajectories is displayed too, for sake of comparison.
The ratio between the actual rate of spreading, and that of the randomized-trajectories,   
is a robust measure for correlations.

We would like to ``see" the correlation by looking on the ``signal".
For this purpose we plot images of the non-randomized sub-sequences in \Fig{fPulseSeq}. 
The sub-sequences are ordered according to their average. 
If the sub-sequences \rmrk{originated} from a randomized-trajectory, 
this average \rmrk{would be} close to zero, and the ordering 
\rmrk{would not result in} any visual effect. 
But sequences of the non-randomized-trajectory are correlated. 
The correlations can be identified by inspection of the figure.
\rmrk{Specifically}, the  sequences of the time-independent  $\theta{=}0$ Hamiltonian  
exhibit long red stretches, and the Kepler-driven sequences exhibit also blue stretches.
This should be contrasted with the sequences of the $\theta{=}\pi$ Hamiltonian,  
that look rather uncorrelated.

\begin{figure}
\includegraphics[width=8cm]{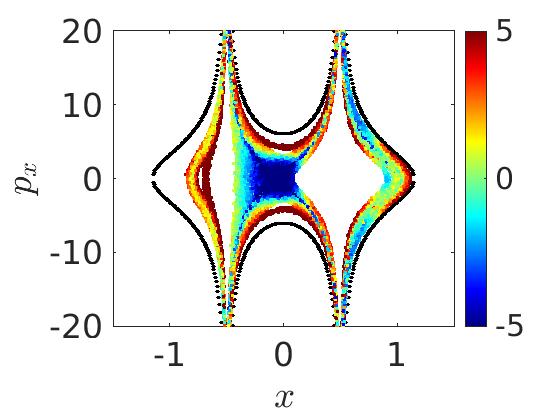}
\includegraphics[width=8cm]{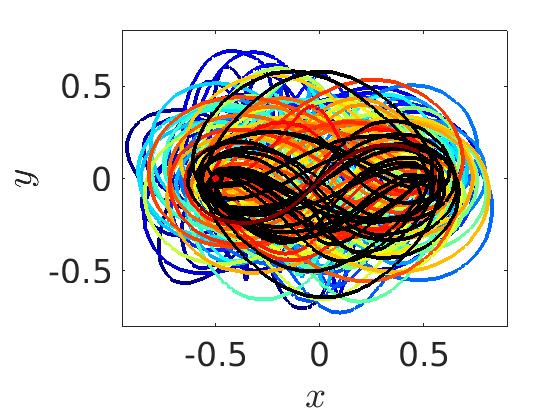} 
\caption{
{\bf Poincaré mapping of pulses.}
The upper panel displays the Poincaré section for the ${\theta=0}$ Hamiltonian. 
Those are the same chaotic trajectories as in \Fig{fLandscape} 
(the quasi-integrable regions are left empty),  
but the points are color-coded  by the values of $\bar{F}_j$. 
The black line indicates the border of the energy surface (the forbidden region is outside).
The lower panel shows a chaotic trajectory $(x(t),y(t))$. 
The color encodes the time. 
The black stretch indicates motion in the ``red" sticky region of the upper panel.
It is characterized by an "$\infty$" shaped loops.        
} 
\label{fPoincare}  
\end{figure}

\begin{figure}
\includegraphics[width=8cm]{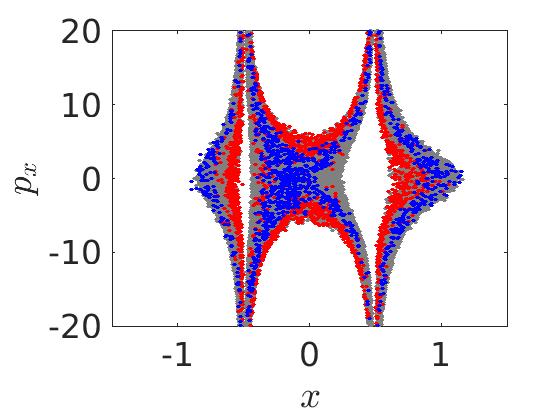}
\includegraphics[width=8cm]{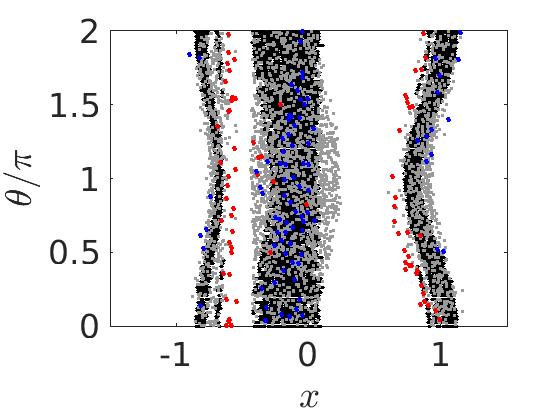}  
\caption{
{\bf Driven system dynamics.} 
In the upper panel a trajectory of the Kepler-driven Hamiltonian
is presented using the same section as that of \Fig{fPoincare}.  
Gray color is used for all the points of the trajectory,  
while red and blue colors are used 
for the {\em correlated stretches} of \Fig{fPulseSeq}.   
Parameters are the same as in \Fig{fLandscape}. 
%
The Hamiltonian is time dependent.  
Accordingly, the cloud of trajectories spreads in energy, 
and can migrate between different regions. 
This is demonstrated in the lower panel, 
where we use the same presentation method as in \Fig{fLandscape}. 
The native chaotic region, that corresponds to the red areas in \Fig{fLandscape}, 
is colored in black.  The evolving cloud of the Kepler-driven system is displayed 
using blue, red and gray dots. Red color indicates sticking in periphery regions, 
while blue indicates sticking in the native chaotic sea. The non-sticking gray points 
expand into ``swamp" regions that are located outside of the native chaotic sea.    
See text for further details. 
} 
\label{fig:cloud}  
\end{figure}

\sect{Phase space exploration}
Having identified correlations in the `signal', 
we would like to trace their phase space origin. 
For this purpose we point out that the value $F$ of the pulse 
provides information about the location of the phase space region that supports the pulse,
as demonstrated in \Fig{fPoincare}. 
Roughly speaking, we can regard $F$ as a radial coordinate 
for points in the Poincaré section. Variations in the value of~$F$ 
indicate migration of the trajectory between different regions. 
In this specific example, 
red pulses originate from peripheral regions of the chaotic sea, 
while blue pulses originate from the central region of the chaotic sea. 
Thus, the red and blue stretches in \Fig{fPulseSeq} indicate {\em stickiness} 
in phase-space regions that have distinct typical non-zero value of~$F$.

The stickiness to peripheral regions is expected.
It has been studied in past literature. 
What we find rather surprising is the extra stickiness that we find in the dynamics that 
is generated by the Kepler driven system: 
the additional {\em blue stretches} indicate excess dwell time in the central region of the chaotic sea.

A more careful inspection, see \Fig{fig:cloud}, 
reveals that the stickiness in the central region of the chaotic sea 
is in the region that was chaotic also in the absence of driving.
So roughly we have the following regions: 
(a)~Native chaotic sea region; 
(b)~Swamp chaotic region; 
(c)~Peripheral chaotic regions; 
(d)~Quasi regular regions. 
The swamp regions appear due to the driving. 
They form in some sense a barrier 
between the native chaotic sea and its periphery. 
As for the quasi regular regions: they are excluded from our simulation, 
and not penetrated by the chaotic trajectories.

\section{Stochastic modelling} 
\label{sec:model1}

Hard-chaos dynamics can be described as a random hopping between cells in phase space. 
We have mixed-chaotic phase space, with tendency for stickiness in e.g. peripheral regions, 
and therefore an effective stochastic description becomes a challenge. 
We would like to introduce a robust procedure for this purpose. 
First of all we recall that: 
(i)~chaotic motion is ergodic. 
(ii)~the pulse strength $F$ is like a radial coordinate. 
It is therefore rather natural to divide phase space into $F$-cells. 
The size of the $F$-bins is determined such that all the (binned) values have 
the same rate of occurrence in the $F_j$ sequence. 
In particular we distinguish in \Fig{fPoincare} 
the blue and the red regions, that corresponds 
to the bins that contain the smallest and the largest pulses respectively.

\sect{Stochastic Kernel} 
Having done the $F$-binning of phase space regions, it becomes possible to define 
a matrix $\bm{P}$ whose element $P_{n,m}$ provide the probability to make a transition form bin~$m$ to bin~$n$.                
Note that the calculation of $\bm{P}$ is a straightforward `signal analysis' procedure 
that is based solely on the inspection of the $F_j$ sequence.
  
\begin{figure}
\includegraphics[width=8cm]{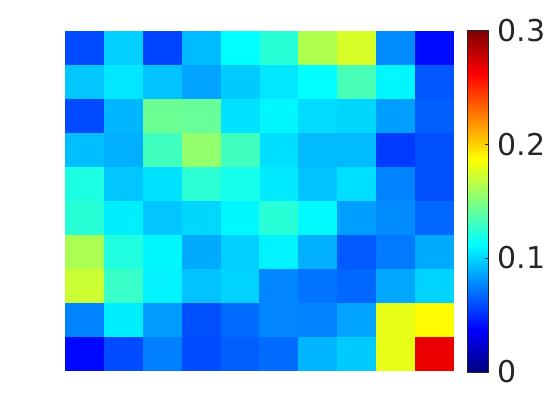} 
\includegraphics[width=8cm]{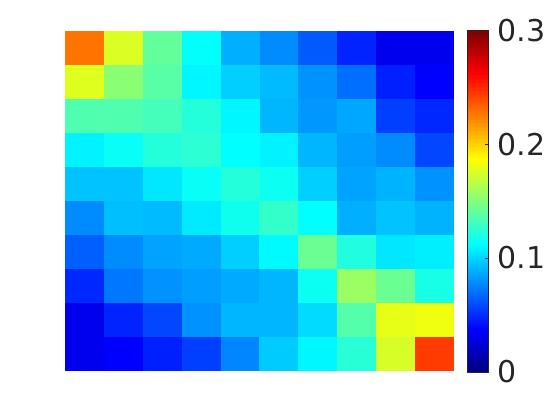} 
\caption{
{\bf Transition probability matrix.} 
The matrix element $P_{n,m}$ is the probability to make a transition form bin~$m$ to bin~$n$.
The images of the matrix are for the frozen ${\theta=0}$ dynamics,
and for the Kepler-driven dynamics. 
In both cases one observes high probability to stay in the red region (last bin). 
In the Kepler-driven case there is also enhance 
probability to stay in the blue region (first bin).
} 
\label{fig:Pmatrix}  
\end{figure}

An image of the $P_{n,m}$ matrix is provided in \Fig{fig:Pmatrix}. 
Qualitatively, we see that the images reflects our 
expectation for enhanced probability to stay in red and blue regions whenever 
stickiness is observed in \Fig{fPulseSeq}. But this is misleading. 
In fact $\bm{P}$ is not capable of providing an explanation for the stickiness. 
We explain this point in the subsequent paragraph.  

\begin{figure}
\includegraphics[width=8cm]{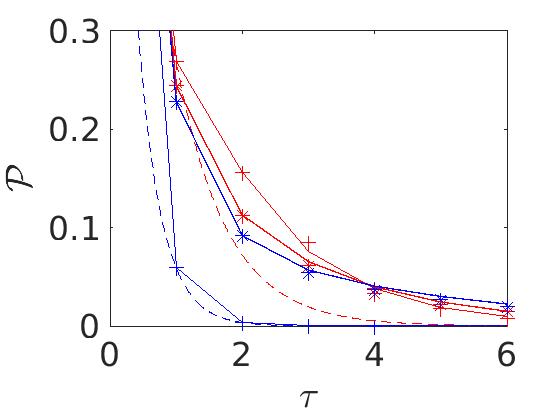} 
\caption{
{\bf Survival probability $\mathcal{P}(\tau)$.} 
The probability to find a pulse that has at least $\tau$ consecutive pulses at the same region. 
The red and blue colors refer respectively to survival in the red and in the blue regions. 
The $+$ symbols are for the time-independent $\theta{=}0$ Hamiltonian, 
and the $*$ symbols are for the Kepler-driven time dependent Hamiltonian.
The solid lines are the analytic results based on the minimal stochastic model
(the thicker lines are for the Kepler-driven Hamiltonian).
The dashed lines illustrate the naive exponential decay 
for the time-independent $\theta{=}0$ Hamiltonian.
}   
\label{fig:Ptau}  
\end{figure}

We can generate artificial $F_j$ sequences using $\bm{P}$ as the propagator (kernel)
for a memory-less Markov process. Naively, one might have the hope to get sequences that 
have the same statistical properties as the original Poincaré sequences. 
But this is not the case: the Markov process does not reproduce the red/blue stretches 
that are seen in \Fig{fPulseSeq}. 
On the quantitative side we define the probability $\mathcal{P}(\tau)$ 
for survival in (say) the ``red" region after~$\tau$ steps.
It is defined as the relative number of ``red" pulses 
that have at least $\tau$ consecutive red pulses. 
(In other words, it is the inverse-cumulative distribution 
of the dwell time in the the red region).  
The one-step survival probability is ${ P_s \equiv \mathcal{P}(\tau{=}1) }$.   
Accordingly, for a Markov process with $P_{n,m}$ we get
\beq \label{eNaive}
\mathcal{P}(\tau)\Big|_{\text{Naive}} \  = \ (P_s)^{\tau}, 
\ \ \ \ \ \ \tau=0,1,2,...  
\eeq
The actual $\mathcal{P}(\tau)$ clearly does not agree with  
exponential decay, as shown in \Fig{fig:Ptau}.  
The naive expectation grossly underestimates the stickiness.

\section{Minimal stochastic model} 
\label{sec:model2}

The failure of $P_{n,m}$ to reproduce $\mathcal{P}(\tau)$ is easily understood by inspection 
of phase space. For presentation purpose we focus on the stickiness in the ``red" region(s)  
of \Fig{fPoincare}. Regarding this region as composed of tiny phase space cells,  
we can determine what is the ``survival time" in the red region for each cell. 
\rmrk{Then we realize that red cells with large survival time constitute a minority}.   
Accordingly, $P_{n,m}$ should be regarded as the coarse-graining of a finer kernel $P_{\nu,\mu}$. 
Note that the type of index (Roman vs Greek) is used in order to distinguish the coarse-grained 
version from the ``microscopic" version.

We would like to construct a {\em minimal} version for $P_{\nu,\mu}$,  
that corresponds to $P_{n,m}$,    
\rmrk{such that  $\mathcal{P}(\tau)$ is reproduced correctly.}
Using this model we would like \rmrk{to relate the rate of spreading to the stickiness}.     
The calculation of $\mathcal{P}(\tau)$, given a Markov kernel $\bm{P}$, 
is done as follows:
\beq \label{ePt}
\mathcal{P}(\tau) \Big|_{\text{Markov}} \ \ = \ \  \left(\frac{1}{\bm{p}^{\dag}\bm{p}}\right) \bm{p}^{\dag} (\bm{Q}\bm{P})^{\tau} \bm{p} 
\eeq 
In this expression $\bm{p}$ is a vector that contains the initial distribution within the bins.
Specifically, we assume uniform distribution within the ``red" bins. 
Note that $(\bm{p}^{\dag}\bm{p})^{-1}$ is the number of participating red bins.
The matrix $\bm{Q}$ is a projector on the red bins, 
and the final projection provides the total survival probability after~$\tau$ iterations with~$\bm{Q}\bm{P}$. 
We can adopt the area of $\mathcal{P}(\tau)$ as a measure for stickiness.
For a given Markov process it is calculated as follows:
\beq \label{eq:stickiness}
\mathcal{S} = \sum_{\tau} \mathcal{P}(\tau) 
=  \left(\frac{1}{\bm{p}^{\dag}\bm{p}}\right) \bm{p}^{\dag} \left[\frac{1}{1-\bm{Q}\bm{P}}\right] \bm{p}
\ \ \ \ \ \ \ 
\eeq
Note that the naive expression \Eq{eNaive} gives 
a rather low value ${ \mathcal{S}=1/(1-P_s) }$.

We already saw in \Fig{fig:Ptau} that a naive 2-region model is not enough to reproduce the stickiness. 
So the minimum is apparently 3-regions.
We assume that we have $n_0$ phase space cells in the non-red region,
and $n_1+n_2$ cells in the red region. 
Consider the possibility of fully-connected chaos 
with ${N=n_0+n_1+n_2}$ cells 
with equal transition probabilities.  
Then the reduced $\bm{P}$ that describes the 
transitions of probabilities between the 3 regions would be 
\beq
\bm{P} = 
\frac{1}{N}
\begin{pmatrix}
	n_0	& n_0 & n_0 \\
	n_1	& n_1 & n_1 \\
	n_2 & n_2 & n_2	
\end{pmatrix}
\eeq
The survival probability in the red region
would be ${P_0=(n_1+n_2)/N}$, 
that is characterized by ${ \mathcal{S}_0=1/(1-P_0) }$.

We now turn to consider a mixed phase-space, where the 
chaotic sea is connected, but not fully-connected. 
Specifically, we assume that the transition probabilities 
from the $n_0$ cells to the $n_1$ cells are 
all equal to $p_1$, while all the transition probabilities 
between the $n_1$ cells and the $n_2$ cells equal $q$.
Then the {\em reduced} $\bm{P}$ that describes the 
transitions of probabilities between the 3 regions is
\beq \label{eq:Pnm}
\bm{P} = 
\begin{pmatrix}
	1-n_1p_1 & n_0p_1 			& 0 \\
	n_1p_1 			& 1-n_0p_1-n_2q 	& n_1 q \\
	0 			& n_2 q 				& 1-n_1 q	
\end{pmatrix}
\eeq
The model is characterized by 4 parameters
\beq \label{eq:P}
P_0  \ &=& \ \frac{(n_1+n_2)}{n_0+(n_1+n_2)} \ \ \equiv \ \ \frac{N_{\text{red}}}{N} \\
\rmrk{R_s} \ &=& \ \frac{n_2}{n_1} \\
\rmrk{P_s} \ &=&  \  1 - \left(\frac{n_1}{n_1+n_2}\right) n_0p_1 \\
\rmrk{Q_s} \ &=&  \ n_1 q
\eeq
The parameters $P_0$ and $R_s$ reflect the relative size of the regions. 
Namely, $P_0$ is the relative size of the red region 
(and hence would equal the survival probability in the red region if we had fully connected chaos), 
and $R_s$ is fraction of sticky red cells. 
The parameters $P_s$ and $Q_s$ reflect the transitions between the regions. 
We could have added also direct transitions 
with probability $p_2$ between the $n_0$ and the $n_2$ cells, 
but it turns out that this would be a redundancy for our purpose.  
Also the total number of cells is insignificant for the analysis.

All the probabilities in the $\bm{P}$ matrix must be less than~$1$. 
This imposes some constraints over the valid range of the model parameters.
In particular one realizes that if ${R_s>1}$ then ${P_s>(1/2)}$. 
Therefore, in order to describe a model that exhibits stickiness (small~$P_s$)
we have to assume ${R_s<1}$, and then the same constraints 
imply that ${ P_s > [R_s/(R_s+1)] }$.   
%
%
%
%

\rmrk{In practice the effective parameters ${(P_s, R_s, Q_s)}$ 
are determined from $\mathcal{P}(\tau)$, as explained in \App{app:S}. 
These parameters determine the stickiness measure of \Eq{eq:stickiness}, 
namely,}  
\beq \label{eq:S}
\mathcal{S} \ \ = \ \ \frac{1}{1{-}P_s} + \left[\frac{R_s}{R_s+1}\right] \frac{1}{Q_s}
\eeq
%
For ${R_s=0}$ we get the naive result ${ \mathcal{S}=1/(1-P_s) }$. 
In the limit ${Q_s\rightarrow 0}$ there is no decay from the $n_2$ cells,
and then the survival probability approaches  
$\mathcal{P}(\infty) = R_s/(1 {+} R_s)$,  
and $\mathcal{S}$ diverges.   
The minimal value ${\mathcal{S}=\mathcal{S}_0 = 1/(1{-}P_0)}$ 
is obtained for a fully connected chaos.

On the basis of the simulations, 
we have determined $\mathcal{P}(\tau)$ 
for the blue and for the red regions, 
for both the $\theta{=}0$ Hamiltonian 
and for the Kepler-driven Hamiltonian. 
The results are displayed in \Fig{fig:Ptau}. 
The effective model parameters have been extracted 
and are listed in \App{app:S}.

\section{Stickiness and rate of spreading} 
\label{sec:SnE}

\rmrk{We would like to relate the rate of spreading to the stickiness.
Within the framework of the stochastic picture,} 
the spreading is determined by the time dependent diffusion coefficient  
\beq
D(t) \ \equiv  \  \frac{1}{2} \sum_{\tau=-t}^t C(\tau) 
\eeq
with the correlation function 
\beq \label{eC}
C(\tau) \ = \ \braket{F_{\tau} F_0} 
\ = \ \frac{1}{N} \sum_{\mu,\nu} f_{\mu} [\bm{P}^{\tau}]_{\mu,\nu}  f_{\nu} 
\eeq
where $N$ is the number of cells, 
and $f_{\mu}$ is the $F$ value that is associated with the phase space cell 
that is indexed as~$\mu$.  
Note that a finite result for $\mathcal{P}(\tau)$ 
is obtained provided ${\sum f_{\mu}=0}$, 
reflecting that the correlation function is defined 
after subtraction of the average \rmrk{(i.e. for a zero average signal)}.

Without the sticky red region, the correlation function 
is of the form $C(\tau)=C_0\delta_{\tau,0}$. 
With the red region included, we get correlations 
due the stickiness. To evaluate the contribution 
of the latter we use the {\em reduced} matrix of \Eq{eq:Pnm},
where all the cells are grouped into 3 regions. 
In such case the sum over $\mu$ in \Eq{eC} is replaced 
by a sum over regions, and the number of cells in each 
region should be introduced as a weight factor.  
Then one obtains 
\beq \label{eC0}
C(0) = C_0 + P_0 \mathcal{S}_0 f_{\text{red}}^2
\eeq 
where $f_{\text{red}}$ is the average $F$ value of the red region, 
and it is implied that the average $F$ value of the non-red region  
is ${ -P_0 \mathcal{S}_0 f_{\text{red}} }$.
The summation over $\tau$ leads to \Eq{eC}
with ${\bm{P}^{\tau}}$ replaced by ${1/(1-\bm{P})}$. 
The zero mode has to be excluded from the inversion.
Including $C_0$ we get 
\beq
\sum_{\tau=0}^{\infty} C(\tau) \ \ = \ \  C_0 \ + \ P_0 \mathcal{S} f_{\text{red}}^2
\eeq
We define the correlation factor as follows:
\beq
\rmrk{c_s}  \ = \ \frac{ \sum_{\tau} C(\tau) }{ C(0) }  
\ = \ 2 \left(\frac{  \sum_{\tau=0}^{\infty} C(\tau) }{ C(0) } \right) - 1
\ \ \ \ 
\eeq
It is the correlation ``time" in terms of iterations with the Poincaré map. 
For the minimal model of \Eq{eq:Pnm} one obtains   
\beq \label{eq:c}
\rmrk{c_s}  \ = \ 2 \left(\frac{\mathcal{S}}{\mathcal{S}_0}\right) - 1, 
\ \ \ \ \ \text{for $C_0{=}0$}
\eeq
Note that for fully connected chaos we get $\rmrk{c_s}=1$ as expected.

\begin{figure}
\includegraphics[width=8cm]{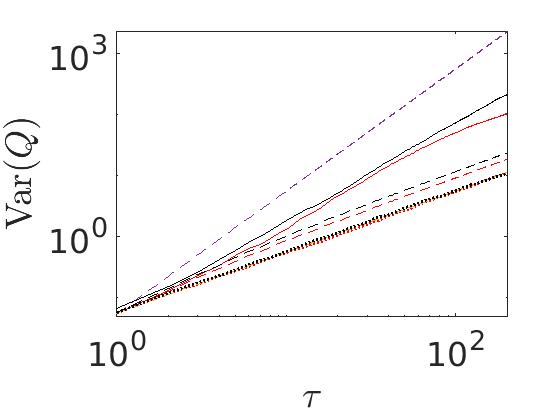}   
\caption{
{\bf Stochastic modelling of \rmrk{spreading}.}
The solid lines are $\text{Var}(Q)$ for the $F_j$ digitized sequences. 
See \App{app:S} for technical details regarding the `digitization'. 
Note that the ``time" ($\tau)$ is the number of Poincaré steps.
The red lines are for the  $\theta{=}0$ Hamiltonian, 
and the black lines are for the Kepler-driven Hamiltonian.
The lower dashed lines are the prediction of the effective stochastic model. 
The dotted lines are for a randomized $F_j$ sequence,
namely, the expected result if the pulses were uncorrelated.   
The lower and upper dashed lines indicate $\propto t$  and $\propto t^2$ dependence
(the former cannot be resolved from the dotted lines).
} 
\label{modelSpreading}
\end{figure}

For the Kepler-driven system we observed an additional ``blue" sticky region. 
Therefore we have to \rmrk{generalize} the minimal model,
such as to have two sticky regions, ``red" and ``blue". 
The total number of cells is ${ N=N_0+N_{\text{red}}+N_{\text{blue}} }$.  
We define ${P_0^{\text{red}}=N_{\text{red}}/N}$ 
and  ${P_0^{\text{blue}}=N_{\text{blue}}/N}$. 
Each of the regions has its own $\mathcal{P}(\tau)$, 
with effective parameters that have been determined in \App{app:S}.
The implied average $F$ value of the $N_0$ region is 
${f_0 = -(1/N_0)[N_{\text{red}} f_{\text{red}} + N_{\text{blue}} f_{\text{blue}}] }$. 
Then one obtains 
\beq
C(0) = C_0 
+ \frac{N_0}{N} f_0^2
+ P_0^{\text{red}} f_{\text{red}}^2 
+ P_0^{\text{blue}} f_{\text{blue}}^2 
\ \ \ \ \ 
\eeq
and
\beq
\sum_{\tau=0}^{\infty} C(\tau) \ = \ C_0 
+ P_0^{\text{red}} \mathcal{S}^{\text{red}} f_{\text{red}}^2 
+ P_0^{\text{blue}} \mathcal{S}^{\text{blue}} f_{\text{blue}}^2 
\ \ \ \ 
\eeq
For $N_{\text{blue}} = 0$ these equations lead back to \Eq{eq:c}.

The correlation factor can be extract numerically from the $S_Q$ plots of \Fig{fSpreading}.
Namely, it is the ratio between the slope of $S_Q^2$ for the true pulse sequence, 
and that of the randomized sequence (of the same pulses).  
By inspection of \Fig{modelSpreading} we see that the true $S_Q^2$ exhibits 
a super-diffusive transient, indicating long-time correlations 
that are not captured by our simplified model. The agreement with the minimal model 
is {\em qualitative} rather than quantitative. Some extra details about the 
quantitative aspect are provided in \App{app:S}. 

We see that a model that faithfully reproduces $\mathcal{P}(\tau)$ 
is not enough for the determination of $C(\tau)$. 
In principle we could have introduced a more elaborated stochastic model, 
that features a hierarchy of red and blue regions, 
but such an approach has no practical value, 
and does not allow the derivation of analytical results.

\section{Cycle vs Modulation} 
\label{sec:cycle}

\begin{figure}
\includegraphics[width=8cm]{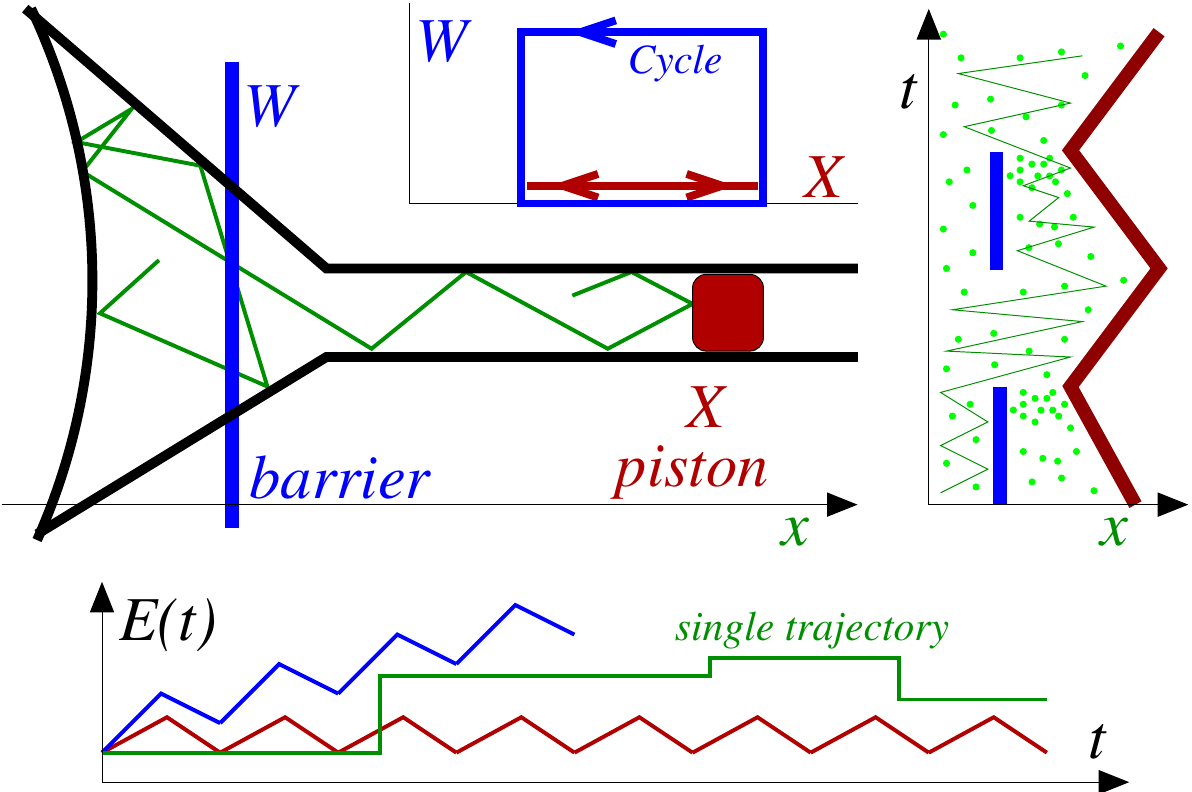}

\caption{
{\bf The billiard paradigm.}
This caricature clarifies the mechanism of dissipation in the quasi-static limit. 
The control parameters are ${\bm{R}=(X,W)}$, 
where $X$ is the position of the piston, 
and $W$ is the height of a dividing barrier. 
Consider the ``blue" quasi-static cycle:
The barrier is turned on ($W=\infty$); 
The piston is pushed in;
The barrier is turned off ($W=0$); 
The piston is pushed out.
Such cycle, unlike the ``red" modulation,  
raises the average energy of the system (${E \mapsto \alpha E}$ with ${\alpha >1}$). 
Note that the blue and red $E(t)$ plots reflect the average over an ensemble of trajectories.  
} 
\label{fBilliard}
\end{figure}

It is important to distinguish between {\em Cycle} and {\em Modulation}. Consider an Hamiltonian $H_{\bm{R}}$ where $\bm{R}$ is a set of control parameters. In a time dependent scenario, we can say that the Hamiltonian varies along a curve in a parametric manifold. A {\em modulation} can be parametrized by a single non-cyclic parameter, say $R(t) = A \cos(\Omega t)$, while a {\em cycle} requires an angle parameter, say ${\theta(t) = \Omega t}$, where $\theta$ is defined modulo $2\pi$. A prototype example for cyclic driving is presented in \Fig{fBilliard}.  

It is sometimes difficult to determine whether the time dependence in the Hamiltonian should be regarded as {\em Constant} or as {\em Modulated} or as {\em Cyclic} driving. For example: the time dependence for particle in a rotating box can be removed by transforming into a rotating frame. Similarly, the time dependence for a particle in an expanding box can be removed via a dilation transformation. In the case of a cycle, the outcome depends in general on the sense of the cycle, and furthermore, for a mixed phase space, we expect difference in the rate of \rmrk{spreading}.   

\begin{figure}
\includegraphics[width=8cm]{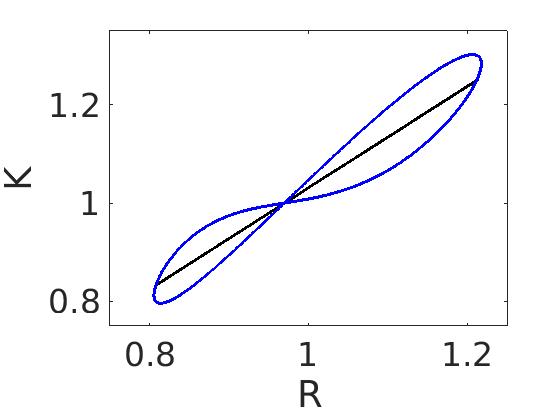}
\caption{
{\bf Cyclic driving.} 
An example for a generic driving cycle in ${(\mathsf{R},\mathsf{K})}$ is illustrated 
by  the blue line, where 
${\mathsf{R}(\theta) \propto  [1 + \varepsilon \cos(\theta) + \varepsilon' \sin(2\theta)]^{-1}}$, 
and $\mathsf{K}$ is determined by \Eq{eq:K}, with $\varepsilon{=}0.2$ and $\varepsilon'{=}0.1$. 
For Kepler driving we set $\varepsilon'{=}0$ and get the black line, 
which is a single-parameter modulation.  
} 
\label{fig:FK}  
\end{figure}

At first glance, one may naively think that a Kepler-driven system qualifies as {\em Cyclic} driving. 
The two parameters might be $(X(t),Y(t))$ or equivalently $(\mathsf{R}(\theta),\mathsf{K}(\theta))$ as in \Eq{eq:H}. 
But it turns out that for a proper Kepler driving the cycle degenerates into a modulation.  
In order to avoid such `degeneracy', we have to assume an asymmetric $\mathsf{R}(\theta)$, 
for example ${\mathsf{R}(\theta) \propto  [1 + \varepsilon \cos(\theta) + \varepsilon' \sin(2\theta)]^{-1}}$, 
that is illustrated in \Fig{fig:FK}.

If we have a non-degenerate cycle, we can ask whether the rate of \rmrk{spreading} depends 
on its {\em sense} (cycle vs reversed cycle). 
For a system with mixed-chaotic phase space indeed we can have such dependence, 
as discussed for e.g. the mushroom billiard in \cite{Kedar3}. 
A different illustration of the same idea is provided in \Fig{fBilliard}.  
During the cycle the space is divided by a barrier (that serves as a ``valve") into two regions. 
This is done periodically, and out-of-phase with respect to the piston movement.
Specifically, in the plotted illustration, the splitting ratio of the cloud 
is roughly 1:2 for forward cycle, and roughly 1:1 for reversed cycle.  
Changes of energy due to changes in the volume obey 
a simple ``ideal gas" multiplicative law ${E \mapsto \alpha E }$, 
where $\alpha$ is given by \Eq{eQa}.
The value of $\alpha$ depends on the sense of the cycle, 
due to the different splitting ratio, 
and we get ${\alpha=10/9}$ and ${\alpha=9/8}$ respectively.

However, the billiard examples are rather artificial. 
They are based on construction that allows a sharp distinction between regions in (phase)space. 
Generic systems, such as the Hill's Hamiltonian, do not feature dramatic splitting and merging 
of well defined (phase)space regions. 
Consequently, dependence on the sense  of the cycle 
is not a prominent effect, and careful numerical procedure is required to detect it.
This motivates the following discussion of {\em directionality} dependence.

\sect{Directionality} 
The dependence on the sense of the cycle is related to the directionality dependence of a modulation. 
The argument is as follows: A modulation can be encoded 
by a sequence $A\bar{A}A\bar{A}\cdots A\bar{A}\cdots$. 
The inverse modulation is clearly the same sequence. A cycle can be encoded 
as $A\bar{B}A\bar{B}\cdots A\bar{B}\cdots$.  
The reversed cycle $B\bar{A}B\bar{A}\cdots B\bar{A}\cdots$ 
is distinct if the cycle is not degenerated (${ A \neq B }$), 
and provided $A$ and $\bar{A}$ are not characterized by the same spreading rate.          
It is therefore enough to establish dependence on directionality.

\begin{figure}
\includegraphics[width=8cm]{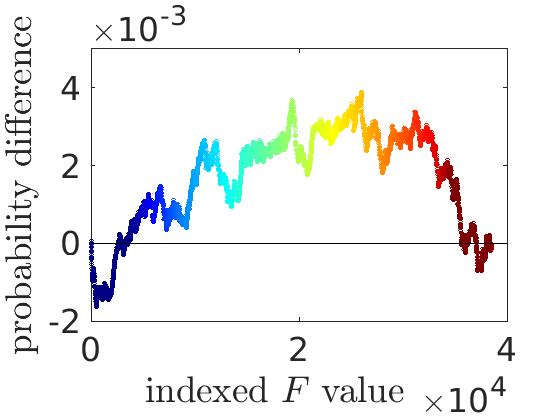} 
\caption{
{\bf Directionality dependence.} 
The probability distribution of the $F_j$ over the bins is uniform by definition. 
It can be regarded as the sum of `forward' pulses distribution, and `backward' pulses distribution.    
Here we plot the deviation of the cumulative probability distribution 
of the backward pulses, from a uniform distribution.
The horizontal axis is the indexed $F$ value (smallest value is indexed as 1, largest value is indexed as 38516).
We see that the small values (blue pulses) become slightly more frequent, 
as opposed to the large values (red pulses) that become slightly less frequent.    
} 
\label{fig:fdistribution}  
\end{figure}

Regarding the sequence $F_j$ as a `signal', we ask whether it looks statistically the same during 
the `forward' half period when $R(\theta)$ changes from $R(0)$ to $R(\pi)$, 
and the `backward' half period when it changes from $R(\pi)$ to $R(0)$.
In the standard paradigm of quasi-static processes the directionality has no significance. 
In \Fig{fig:fdistribution} we plot the distribution of the $F$ values for the two groups of pulses. 
For the full signal the distribution of the $F_j$ over the bin is uniform by definition. 
But if we look only on the pulses that belong to the `backward' half-periods 
we see that the small values (blue pulses) become slightly more frequent, 
as opposed to the large values (red pulses) that become slightly less frequent. 
The difference is very small.  Still, it indicates that the steady state 
is not the same for ``forward" and ``backward" driving.

\section{Dissipation} 
\label{sec:dissipation}

Dissipation is associated with energy spreading.   
The standard theory \cite{Ott1,Ott2,Ott3} 
assumes a globally chaotic energy surface that instantly ergodizes at any moment.  
It follows that the phase-space volume $\mathcal{N}(E;\theta)$ is an adiabatic invariant, 
where $\theta(t)$ is a slowly varying control parameter.
For a closed cycle, the conservative work is zero. Still, beyond the zero-order adiabatic result, 
there is diffusion in energy with coefficient ${D_E = \nu \dot{\theta}^2}$,  
where $\nu$ is the intensity of the fluctuations, i.e. the algebraic area of $C(t)$. 
From the Fokker-Plank description of the spreading process, 
one deduces the rate of absorption ${\dot{E} = \mu \dot{\theta}^2}$,
aka the Kubo formula, with dissipation coefficient $\mu$ that is related to $\nu$ 
via a fluctuation-dissipation relation \cite{Wilkinson1,Wilkinson2,crs,frc}, 
namely ${\mu = (1/2)\beta \nu}$, where $\beta(E)$ is some version of 
microcanonical inverse temperature, as defined in \App{app:qes}.
Consequently, for periodic driving with frequency ${\Omega \equiv \dot{\theta}}$  
one expects an amount ${ Q = 2\pi \mu \Omega }$ of dissipated energy per cycle,   
which vanishes in the quasi-static limit (${\Omega \rightarrow 0}$).

For a driven mixed-chaotic system, we expect parametric dissipation, 
meaning that the dissipated energy per cycle (${ Q }$) 
approaches a finite non-zero constant in the limit $(\Omega \rightarrow 0)$, 
and depends on the directionality of the driving as discussed in the previous paragraph. 
Billiard examples that have been discussed in the past, 
as well as that of \Fig{fBilliard}, are illuminating, 
but do not fully reflect some complications that are 
encountered once we deal with a generic system, such as Hill's.   
In what follows we highlight those zero-order subtleties, 
and also generalize the first-order formulation.

\sect{Zero order dissipation}
In systems with mixed-chaotic dynamics, we can get irreversibility 
due to phase space spreading (\rmrk{aka growth of the entropy}), 
as well as dissipation (growth of the average energy), even in the quasi-static limit.
%
%
The derivation of this claim requires phase-space generalization of \cite{Kedar3}. 
This generalization is presented in \App{app:qes}.
We write the phase space area as ${\mathcal{A} = \sum_{\mu}  \mathcal{A}_{\mu}}$, 
where ${\mu}$ distinguishes different regions. 
Each region might have a different ``inverse temperature" ${\beta_{\mu} }$. 
Then we obtain the following result: 
\beq \label{eQ}
\braket{Q}_{0} \ \ &\approx& \ \ 
- \sum_{steps} \sum_{\mu}   \frac{\delta \mathcal{A}_{\mu}}{\beta_{\mu} \mathcal{A}} 
+  \mathcal{O}(\delta \mathcal{A}^2)
\eeq 
This expression is obtained from \Eq{eQb} after expansion 
with respect to ${\delta \mathcal{A} = \mathcal{A}_{\mu}-\mathcal{A}_{\mu}^{(0)}}$.
Based on \Eq{eQ} our observation is that we can get an $\mathcal{O}(\delta \mathcal{A})$ non-zero result provided the $\beta_{\mu}$ are non-identical. 
In such case $\braket{Q}_{0}$ can switch sign for a reversed cycle. 
This should be contrasted with a Billiard system for which the ${\beta_{\mu}=1/E}$ are identical, 
and $\braket{Q}_{0}$ is always positive.

\sect{First order Dissipation}
We define ${ \mathcal{F} = -\partial \mathcal{H}/\partial \theta }$.
For a periodically driven Hamiltonian with ${\dot{\theta} = \Omega }$ 
we have ${ \dot{E} = -\Omega  \mathcal{F}(t) }$. 
Integrating over a cycle, squaring, and averaging over an ensemble, 
we get ${\text{Var}(Q)  =  2 \pi \nu \Omega }$, 
where $\nu$ is the intensity of the fluctuations 
(the area of the $\mathcal{F}$ autocorrelation function). 
This assumes a globally chaotic energy surface. 
If we have a fragmented phase-space (as in the previous Billiard example) 
we get ${\text{Var}(Q)  =  \text{Var}_0(Q) + 2 \pi \nu \Omega }$,  
where $\text{Var}_0(Q)$ is the variance 
that is associated with $\braket{Q}_{0}$. 
Using the above explained fluctuation-dissipation reasoning  
we deduce that the energy increase per cycle is 
\beq \label{eQcycle}
\braket{Q} \ \ = \ \ \braket{Q}_{0} 
\ + \  \frac{1}{2} \beta_{\text{eff}} \left[ \text{Var}_0(Q) + 2\pi \nu \Omega \right] 
\eeq
This expression goes beyond Kubo, because the zero order spreading is taken into account. 
Note that if correlations persist over a time duration that is longer than a cycle, 
the result is a long super-diffusive transient as in \Fig{fSpreading}.  
In any case the appropriate correlation factor $c_s$ has to be incorporated 
in the calculation of~${\nu}$, as discussed in \Sec{sec:SnE}.

\section{Summary and outlook}

We have introduced an effective stochastic theory for quasi-static \rmrk{spreading} in systems with mixed chaotic phase space. \rmrk{The main objective was to provide tools for the analysis of phase space spreading.  More specifically, the spreading of the energy, which is useful for the calculation of the average energy growth (dissipation), and possibly for estimating the rate of ``evaporation".} 

\rmrk{For demonstration of our approach we have selected Hill's Hamiltonian. This toy model, by itself, has physical significance, as discussed in the Introduction. The problem of interest has possibly direct relevance to studies that concern the long-term stability of planets in binary systems \cite{hill7,hill8,hill9}. Furthermore, it illuminates the relevance of mixed-chaotic dynamics in the context of the 3~body problem.  The model has all the essential ingredients for our analysis. However, in retrospective, we have to admit that {\em stickiness}, rather than a {\em zero-order dissipation} effect, is the dominant feature that determines the rate of spreading. This stands in contrast to the analysis of energy spreading due to quasi-static driving of specially-designed Billiard systems \cite{Kedar1,Kedar2,Kedar3}, that has motivated the present study.}             

Our agenda was, on the one hand, to characterize the multi-dimensional phase space dynamics via ``signal-analysis" of a single chaotic trajectory. On the other hand, we wanted to reproduce the essential statistical features of the `signal' using a minimal Markovian model. 

For the characterization of the chaotic motion, we represent the chaotic trajectory as a Poincaré sequence of pulses~($F_j$). The value of $F$ is regarded as a `radial' phase space coordinate, that is used in order to divide phase space into {\em regions} (indexed by $n$).          
We realize that this coarse-graining is too rough: we cannot build on it a Markov process 
that reproduces the observed stickiness. We therefore have to define a refined version 
of the Markov process that reflects the hierarchic structure of phase space. 
Consequently, we constructed a minimal model that allows to reproduce the observed stickiness.
This model suggests a relation between the stickiness and the enhancement 
that is observed is the rate of \rmrk{spreading}. Unfortunately, in the present model, 
the quantitative agreement is poor due to long range correlations that were neglected.

Specifically, for the frozen dynamics, we have identified stickiness in peripheral regions of the chaotic sea. A minimal stochastic model for such configuration requires 3 regions (central chaotic region;  non-sticky peripheral chaotic region; sticky peripheral chaotic region).  Surprisingly in the Kepler-driven system extra stickiness manifests in the native chaotic sea. This extra stickiness is related to the appearance of an additional ``swamp chaotic region", where chaos penetrates due to the time-dependence of the Hamiltonian. Nevertheless, it can be treated on equal footing using the same stochastic model (with extra regions).  

We also looked for directionality dependence, implying that the rate of \rmrk{spreading} is not the same if a cycle is reversed. We have clarified that also this effect can be identified from the ``signal analysis" of the Poincaré sequence. For the model system that we have studied, the finding was that it is a very weak effect (a few percent difference).  

Finally, for sake of generality, we have explained how the Kubo theory of dissipation   
can be generalized in order to incorporate both the zero-order and first-order irreversibility.
This picture implies exponential energy growth if $\braket{Q}$ of \Eq{eQcycle} is proportion to~$E$.  
This is indeed the case for Billiard systems if ${\braket{Q}_0 \ne 0}$ 
as discussed originally by \cite{Kedar1,Kedar2,Kedar3}.  
More generally we can get from \Eq{eQcycle} different energy dependence, 
say ${ \dot{E} = \lambda E^{\alpha} }$.
Note that for ${ \alpha >1 }$ one obtains hyperbolic-like growth that leads 
to escape ${E \sim 1/(t-t_e)^{1/(\alpha-1)}}$ within a finite time~$t_e$. 
The exploration of such scenario requires further study of possibly different model systems.

\appendix

\section{Basic formulas for Kepler motion} 
\label{app:kepler}

The constant of motion in Kepler problem is the angular momentum.
In terms of polar coordinates $(\theta,R)$ we define ${\ell = R^2 \dot{\theta}}$. 
Kepler's area law is the statement
\beq
\frac{d}{dt}\text{Area} = \frac{1}{2}\ell 
\eeq 
The Kepler motion is along an ellipse with major axes  ${ a }$ and ${ b = \sqrt{1-\varepsilon^2} a}$.
We also define ${c=\sqrt{ab}}$. From the area law it follows that ${T = 2\pi ab /\ell }$. 
Accordingly the frequency is ${\Omega = \ell/(ab)}$. So we have the relation 
\beq \label{eT}
\ell \ \ = \ \ ab\Omega  \ \ = \ \ c^2 \Omega
\eeq 
For a circular motion of radius $R{=}a{=}b{=}c$, 
the frequency the motion is determined by the equation 
\beq \label{eO}
\Omega^2 a^3 = G(M_1+M_2) \equiv GM
\eeq
This result applies also if the motion is along an ellipse.
The equation of the ellipse is 
\beq
R(\theta) \ \ = \ \ \frac{(1-\varepsilon^2) \, a}{1 + \varepsilon \cos(\theta)}  \ \ \equiv \ \ c \mathsf{R}(\theta)
\eeq
Note that with this definition $\mathsf{R}(\theta)$ is square-normalized to unity.  
The equation of motion for the radial motion is 
\beq
\ddot{R} = \frac{\ell^2}{R^3} - \frac{GM}{R^2} 
\eeq
which implies conservation of energy (here we are in the non-rotating ``lab" frame):  
\beq \label{eE}
E \ = \ \frac{1}{2}\dot{R}^2 + \frac{\ell^2}{2R^2} - \frac{GM}{R} 
\ = \ -\frac{1}{2} a^2 \Omega^2  
\eeq

Given $(GM, E, \ell)$, the orbit, up to orientation,  
is described by $(\Omega,a,c)$.
The $\Omega$ and the $a$ are determined by \Eq{eO} and \Eq{eE},
while $c$ is determined by \Eq{eT}, 
and we have the ratio ${c/a=(1-\varepsilon^2)^{1/4} }$.

\hide{  
\beq
\ell \ \ = \ \ \frac{GM}{(-2E)^{1/2}} \sqrt{1-\varepsilon^2}
\eeq
The time dependence of the Kepler motion is found as follows.  
The eccentric angle  $\alpha(t)$ is found via Kepler's equation:  
\beq \label{e28}
\alpha -\varepsilon \sin(\alpha)  = \Omega t 
\eeq
Then the polar angle $\theta(t)$ is found via
\beq
\tan(\theta/2) =  \sqrt{\frac{1-\varepsilon}{1+\varepsilon}}  \tan(\alpha/2)  
\eeq
and the radial coordinate $R(t)$ via 
\beq \label{e30}
R(t) = [1- \varepsilon \cos(\alpha)] a 
\eeq
In simulations it might be convenient to generate $t$ as a function of $\alpha$ via \Eq{e28} and from \Eq{e30} to get~$R(t)$. 
}

\section{The generalized Hill Hamiltonian}
\label{app:hillG}

We use the notations ${\bm{r}=(x,y)}$ and ${\bm{p}=(p_x,p_y)}$.
We consider time dependent $R(t)$ and $\theta(t)$.  
Without loss of generality, we set $\mass {=} 1$ for the mass of the satellite.
The Hamiltonian is:
\beq
\mathcal{H} \ \ = \ \ \frac{1}{2} \bm{p}^2 \ + \ U(\bm{r};R(t),\theta(t))
\eeq
In order to transform the Hamiltonian we use a sequence of canonical transformations. For clarity we use below ``quantum language". Given a transformation $T=\exp[-i \alpha(t) G]$ that is generated by $G$, we use below the formula
\beq
\mathcal{H} \ \ &=& \ \ T^{\dag}\mathcal{H}T - i T^{\dag} \frac{\partial T}{\partial t} 
\\
\ \ &=& \ \ T^{\dag}\mathcal{H}T - \dot{\alpha} G
\eeq
\rmrk{The first transformation is to a rotating reference frame} with ${T = \exp[-i \theta(t) L]}$ where ${L = \bm{r} \wedge \bm{p} = x p_y - x p_y}$
\beq
\mathcal{H} = \frac{1}{2} \bm{p}^2 -  \dot{\theta} L + U(\bm{r};R(t),0)  
\eeq
\rmrk{The second transformation is a time dependent dilation} with ${T = \exp[-i (\ln R) K]}$ where ${ K = \bm{r} \cdot \bm{p} = x p_x + y p_y }$. 
Note that ${T^{\dag}x T = R x}$ and ${T^{\dag}p T = (1/R)p}$.
Using the notation ${U(r) = U(r;1,0)}$ we get   
\beq
\mathcal{H} &=& \frac{1}{2R^2}\bm{p}^2  - \frac{\dot{R}}{R}K  -\dot{\theta}L   + \frac{1}{R}U(\bm{r}) \\ 
&=& \frac{1}{2R^2}\bm{p}^2  - \frac{\dot{R}}{R}\bm{r}\cdot\bm{p}  -\dot{\theta}L   + \frac{1}{R}U(\bm{r}) 
\eeq
\rmrk{The third transformation is a time dependent Gauge} with ${T = \exp[-i \Lambda] }$
where ${\Lambda = - (1/2) R\dot{R} r^2 }$.
Note that ${T^{\dag} x T = x}$ and ${T^{\dag} p T = p - \partial \Lambda }$.
Accordingly we get
\beq \nonumber
\mathcal{H} &=& \frac{1}{2R^2}(\bm{p} + R\dot{R} \bm{r})^2 
- \frac{\dot{R}}{R} \bm{r}\cdot (\bm{p} + R\dot{R} \bm{r}) 
-\dot{\theta}L    
\\ \nonumber 
&& + \frac{1}{R}U(\bm{r}) + \frac{1}{2} [\dot{R}^2 + R\ddot{R}] \bm{r}^2 
\\ \nonumber
&=& \frac{1}{2R^2}\bm{p}^2  - \dot{\theta}L + \frac{1}{R}U(\bm{r}) + \frac{1}{2} R\ddot{R} \bm{r}^2 
\\ \nonumber
&=& \frac{1}{2R^2}(\bm{p}-\bm{A})^2 + \frac{1}{R}U(\bm{r}) -\frac{1}{2}\left[R^2\dot{\theta}^2 - R\ddot{R}  \right] \bm{r}^2 
\eeq
where ${\bm{A} = R^2\dot{\theta} \bm{r}_{\perp}}$ with ${ \bm{r}_{\perp} = (-y,x)}$.
Given $R(t)$ and $\theta(t)$, the above Hamiltonian can be written schematically as 
\beq \nonumber
&& \mathcal{H}(\bm{r},\bm{p}; \, \theta(t), R(t)) \ \ = \ \  
\\ \nonumber
&& \frac{1}{R(t)^2} \left\{ \frac{1}{2}(\bm{p}-\ell(t) \bm{r}_{\perp} )^2 + R(t) U(\bm{r}) - \frac{1}{2} \rmrk{K(t)} \bm{r}^2 \right\} 
\eeq
where $\ell(t) \equiv R^2\dot{\theta}$,
and \rmrk{$K(t) \equiv (R^2\dot{\theta})^2 - R^3\ddot{R}$.}
If we assume Kepler motion we get \rmrk{${K(t) = \Omega^2 a^3 R(t)}$}
from the radial equation of motion.

\section{Hamiltonian for a Kepler system}
\label{app:hillK}

Due to the dilation transformation, the coordinate $r$ is dimensionless, and the distance between the stars is unity, 
while $p$ has the same units as $\ell$. We now assume that $\ell$ is constant for the cycles of interest. 
Consequently we can re-scale the momentum ${p:=\ell p}$.  
It is convenient to define the characteristic radius $c$ of the orbit through ${\ell \equiv \Omega c^2}$, 
where $\Omega$ is the frequency of the cycle. 
We also define the notation $R(t)=c\mathsf{R}(\theta(t))$.  
By definition of $c$ and from  ${\dot{\theta}=\ell/R(t)^2}$ 
it follows that 
\beq
\oint \left| \mathsf{R}(\theta) \right|^2 \, \frac{d\theta}{2\pi} \ \ = \ \ 1
\eeq 
Given $\mathsf{R}(\theta)$ 
we have the identity
\beq
\frac{\ddot{\mathsf{R}}}{\Omega^2} \mathsf{R}^3 \ \ = \ \ - \left(\frac{1}{\mathsf{R}}\right)'' \mathsf{R} 
\eeq
where dot ($^.$) is for time derivative and prime (') is for theta derivative.

We write the attraction constant between the satellite and the stars as $G_0M$, 
such that $U(r)= G_0M u(r)$. 
The Hamiltonian takes the form
\beq 
\mathcal{H} =  \frac{\Omega}{\mathsf{R}^2}
\left\{ \frac{1}{2}(\bm{p} - \bm{r}_{\perp} )^2 + g \mathsf{R} u(\bm{r}) - \frac{1}{2} \mathsf{K} \bm{r}^2 \right\} 
\ \ \ \ \ 
\eeq        
where
\beq
\mathsf{K} \ = \ 1 +  \left(\frac{1}{\mathsf{R}}\right)'' \mathsf{R}
\eeq
and
\beq
g \ = \ \frac{G_0M}{c^3\Omega^2} \ \equiv \ \frac{\Omega_0^2}{\Omega^2}
\eeq
For a Kepler driven system we use the notation 
\beq
g_{\varepsilon} \ = \ \frac{GM}{c^3\Omega^2}  \ = \ (1-\varepsilon^2)^{-3/4}
\eeq
and get the simpler Hamiltonian 
\beq \label{eH} 
\mathcal{H} 
= \frac{\Omega}{\mathsf{R}^2}
\left\{ \frac{1}{2}(\bm{p} - \bm{r}_{\perp} )^2 +  \mathsf{R} \left( g u(\bm{r}) - \frac{1}{2}g_{\varepsilon} \bm{r}^2 \right) \right\} 
\ \ \ \ 
\eeq        
Given $\mathsf{R}(\theta)$ and $\Omega$ and $c$ we have ${\dot{\theta}=\Omega/\mathsf{R}^2}$.               
Consequently, if we use $\theta$ as time variable, 
we get the Hamiltonian \Eq{eH} without the $\Omega/\mathsf{R}^2$ term.

The simple minded slowness condition is ${ \Omega \ll \Omega_0 }$, 
which can be written as ${ 1 \ll g }$.  
In analogy with the piston paradigm, 
we have to assure that ${ \dot{R} \ll \dot{r} }$ 
where the typical velocity of the dust particles is ${\dot{r} \sim c\Omega_0 }$.
For Kepler motion, the maximum velocity of the ``piston" is ${ \dot{R} \sim c \epsilon g_{\epsilon} \Omega }$. Consequently, the slowness condition takes the form ${ \varepsilon g_{\varepsilon} \ll g }$, which always breaks down if ${ \varepsilon }$ is too close to unity.

\section{Determination of effective parameters} 
\label{app:S}

The $F_j$ values have been grouped into 10 bins. The pulses that belong to a given bin define a region $n$ in phase space. It is implied that the same number of pulses is associated with each region. In our jargon ${n=1}$ is the ``blue" region and ${n=10}$ is the ``red" region, \rmrk{and it is implied that for fully connected chaos the probability to stay in a red bin is ${P_0=0.1}$.} The matrix $P_{n,m}$ of \Fig{fig:Pmatrix} characterizes the statistics of the transitions between regions. Additionally, we determine numerically the probability $\mathcal{P}(\tau)$ to stay in a given region as a function of $\tau$, see \Fig{fig:Ptau}. 
\rmrk{Specifically, we have obtained $\mathcal{P}(\tau)$ for $[\theta{=}0,\text{red}]$, for $[\theta{=}0,\text{blue}]$, for $[\text{Kepler,red}]$, and for $[\text{Kepler,blue}]$. From that we have extracted (in each case) the staying probability ${P_s = \mathcal{P}(\tau{=}1)}$, and the stickiness measure $\mathcal{S}$. The latter is  the `area' of $\mathcal{P}(\tau)$. The additional effective parameters $(R_s,Q_s)$ are deduced via \Eq{eq:S}, with value of $1/Q_s$ that fits the stretched tail of $\mathcal{P}(\tau)$. The results were respectively:}    
\beq
P_s &=& 0.27,0.06,0.24,0.23 \\
R_s &=& 0.15,0.0003,0.092,0.067 \\
Q_s &=& 0.65,0.83,0.42,0.26 \\
\mathcal{S} &=& 1.57,1.06,1.52,1.53 
\eeq
The main difference between the Kepler-driven Hamiltonian and the frozen Hamiltonian 
is related to the stickiness in the blue region.

The {\em `digitized'} signal is obtained as follows. 
We define $f_n$ as the average value 
that characterizes the \mbox{$n$-th} bin. 
Then we set ${\text{digitized}[F_j] = f_n}$, 
if $F_j$ belongs to the $n$-th bin.
In order to analyze the stickiness-related correlations, 
we have regarded all the intermediate bins ($n=2 \cdots 9$) 
as one region that is characterized by an average value $f_0$, 
while bins ${n=1,10}$ are characterized by $f_{\text{blue}}$ and $f_{\text{red}}$ respectively.
Due to this digitization the noise is reduced by factor ${\sim 2.5}$. 
We are left with a signal that contains information 
that is related to the stickiness, 
and we can set ${C_0=0}$ in \Eq{eC0}. 
Consequently, this digitization procedure allows a meaningful comparison 
between the numerical results and the minimal model in \Fig{modelSpreading}.

The correlation factor $\rmrk{c_s}$ can be extract numerically 
by inspection of \Fig{modelSpreading}.   
For the Kepler-driven system we get ${\rmrk{c_s} = 18}$,  
while for the $\theta{=}0$ Hamiltonian we get ${\rmrk{c_s} = 9}$. 
This is consistent with what we observed in \Fig{fSpreading}.
The minimal model does not take into account 
the observed long-time correlations, and therefore 
predicts much smaller values, 
namely, ${\rmrk{c_s}  = 2.0}$ and ${\rmrk{c_s} = 1.7}$ respectively.

\section{Quasi-static energy spreading} 
\label{app:qes}

The energy landscape of phase space is described by the function $E=\mathcal{H}(\bm{r},\bm{p};\theta)$.  
The $d\bm{r}d\bm{p}/(2\pi)^{\text{dof}}$ volume of an energy surface is denoted $\mathcal{N}(E;\theta)$, 
and corresponds to the number of phase space cells in semiclassical mechanics.    
The area of the energy surface is defined as ${ \mathcal{A}(E;\theta) = \partial_E \mathcal{N} }$,
and corresponds to the density of states.  
The microcanonical-like inverse temperature is  $\beta = \partial_E \ln \mathcal{N} = \mathcal{A}/\mathcal{N}$. 
For a particle in a billiard  of area $\mathcal{A}$, 
setting appropriate units for the mass, 
we get ${\mathcal{N} = \mathcal{A} E}$, and ${\beta = 1/E}$.
For a mixed phase space the total area is written as   
\beq
\mathcal{A}(E;\theta) \ \equiv \ \partial_E \mathcal{N}  \ = \ \sum_{\mu} \mathcal{A}_{\mu}(\theta)
\eeq
This assumes that there is a way to identify distinct regions as in the billiard example of \Fig{fBilliard}
where ${{\mu}=L,R}$ distinguishes the left and right regions, 
and $\mathcal{A}_{\mu}(\theta)$ is the respective geometric area of the ${\mu}$-th region, 
while $\theta$ is a parameter that is used to specify the position of the piston.  
Without any approximation we always have 
\beq
\dot{E} \ = \ \braket{ \frac {\partial \mathcal{H}} {\partial \theta} }_t \dot{\theta}
\ \equiv \ -  \Omega \mathcal{F}(\theta(t))
\eeq
In the Ott-Wilkinson-Kubo formulation of linear response theory \cite{Ott1,Ott2,Ott3,Wilkinson1,Wilkinson2,crs,frc},  
it is assumed that for a quasi-static process the instantaneous average can be replaced by 
an evolving microcanonical average due to quasi-ergodicity. 
Accordingly, the variation of the energy becomes parameteric:   
\beq
dE \ = \ \braket{ \frac{\partial \mathcal{H}}{\partial \theta} }_{E,\theta} \!\!\!d\theta 
\ \ = \ \  -\left(\frac{\partial_{\theta} \mathcal{N}}{ \partial_E \mathcal{N}}\right) d\theta
\eeq
From the last relation it is implied that ${d\mathcal{N}=0 }$, 
meaning that $\mathcal{N}(E;\theta)$ is an adiabatic invariant. 
With the definition of phase space area this can be written as 
\beq
dE =   -\frac{1}{\beta} \, [\partial_{\theta} \ln \mathcal{N}] \, d\theta 
\equiv -\frac{1}{\beta_{\text{eff}}} \, [\partial_{\theta} \ln \mathcal{A}] \, d\theta 
\eeq
where the latter equality defines $\beta_{\text{eff}}$.
Adjusting notations to mixed phase space we write 
the change of the energy per-cycle as  
\beq
dE \ =  -\sum_{\mu}  P_{\mu}(\theta) \frac{1}{\beta_{\mu}} \, [\partial_{\theta} \ln \mathcal{A}_{\mu}] \, d\theta 
\eeq
where $P_{\mu}(\theta)$ is the probability at region~${\mu}$ of the energy surface, 
and it is assumed that the regions are well defined. 
Ref \cite{Kedar3} consider a more complicated case where the borders between regions 
is affected by~$\theta$. But such complication does not affect the big picture.

For a Billiard system that undergoes a multi-step process 
of the type that is illustrated in \Fig{fBilliard},  
the dissipated energy per cycle is 
\beq \label{eQb}
\braket{Q}_{0} \ \ = \ \ 
- \frac{1}{\beta}  \sum_{steps} \sum_{\mu} 
\frac{\mathcal{A}_{\mu}^{(0)}} {\mathcal{A}^{(0)}} \ln\left[ \frac{\mathcal{A}_{\mu}}{\mathcal{A}_{\mu}^{(0)}}  \right]  
\eeq 
where $\beta=1/E$ assumes a narrow distribution around~$E$.
Here the outer summation is over {\em steps} of the cycle. 
We assume global chaos at transitions between steps.
The superscript "0" indicates the area at the beginning of a step. 
Without  "0" it is the area at the end of the step.

Billiard systems are simple enough to allow an improved (exact) version of \Eq{eQb} 
that does not assume a narrow distribution around a {\em fixed} energy. 
Changes of energy due to changes in the volume obey the simple ``ideal gas" 
multiplicative law ${E \mapsto \alpha E }$, with 
\beq \label{eQa}
\alpha \ \ = \ \ 
\sum_{steps} \sum_{\mu} 
\frac{ (\mathcal{A}_{\mu}^{(0)})^2 } { \mathcal{A}^{(0)} \mathcal{A}_{\mu} }    
\eeq 
One can easily verify that $\braket{Q}_{0}$ of \Eq{eQb} 
is consistent with $(\alpha-1)E$. 
Note that we always have ${\alpha > 1}$.

\ \\

\sect{Acknowledgment}
\rmrk{We thank Hagai Perets and Nicholas Stone for a helpful communication.}  
This research was supported by the Israel Science Foundation (Grant No.283/18).

\clearpage
\end{document}